\documentclass{article}
\usepackage[utf8]{inputenc}
\usepackage{enumitem}
\usepackage{graphicx}
\title{Behavioral Biases and Nonadditive Dynamics in Risk Taking: An Experimental Investigation}
\author{José Cláudio do Nascimento \footnote{ Universidade Federal do Ceará, adress:  Rua Coronel Estanislau Frota, 563 - Bloco I - Centro - Campus de Sobral - Mucambinho - CEP 62010-560 - Sobral - CE, Brazil, email: claudio@sobral.ufc.br} }

\date{}

\begin{document}

\maketitle
\begin{abstract}
This paper investigates the dynamics of gambling and how they can affect risk-taking behavior in regions not explored by Kahneman and Tversky's Prospect Theory. Specifically, it questions why extreme outcomes do not fit the theory and proposes alternative ways to measure prospects. The paper introduces a measure of contrast between gambles and conducts an experiment to test the hypothesis that individuals prospect gambles with nonadditive dynamics differently. The results suggest a strong bias towards certain options, which challenges the predictions of Kahneman and Tversky's theory.
\end{abstract}
\section{Introduction}
Taking risks is an activity that has a direct impact on the health of individuals, which justifies research into the behavior of decision makers \cite{krotava2018psychological, kandasamy2014cortisol, cueva2015cortisol}. Among the works that stood out in this field  is the Prospect Theory. This theory demonstrates that people think in terms of expected utility relative to a reference point (eg, current wealth) rather than absolute outcomes \cite{kahneman2013choices,kai1979prospect}. Among the modeled behaviors, two behaviors stand out for having broad empirical validation:

\begin{enumerate}
    \item risk aversion for positives prospects - people prefer gains with low uncertainty to those gains with high uncertainty, even if the average of uncertain gains is more significant than the certain gain.
    \item risk seeking for negative prospects - people  prefer losses with high uncertainty to those certain losses, even if the average of uncertain losses is lower than the certain loss.
\end{enumerate}
In the paper where the Prospect Theory was first presented, we can note that only one type of average was adopted as a rational expectation of gambles \cite{kai1979prospect}. This average was compared to the experiment and it was shown that it did not describe the behavior of risk takers. Thus, the authors proposed a theory that described the observed behaviors.

In a second version of the Prospect Theory, Kahneman and Tversky highlighted that ``the shape of the weighting functions favors risk seeking for small probabilities of gains and risk aversion for small probabilities of loss, {\it provided the outcomes are not extreme}''\cite{tversky1992advances}. This detail leads us to the following question: Why doesn't the theory work with extreme outcomes? This paper investigates possible dynamics that a gamble can have and how these dynamics can affect the behavior of risk takers in regions where Kahneman and Tversky do not explore.

If Kahneman and Tversky's experiments revealed that prospect deviates from the average of probability-weighted outcomes, then could some other averages be adopted as a prospect by risk-takers heuristic?
In statistics, the average is defined as the value that reveals the concentration of data in a distribution. For its calculation, many repetitions of the event are necessary. However, there are some different types of averages and each one applies to a context of how the results evolve over the  repetitions \cite{peters2016evaluating}. Furthermore, repetitions are not just a mathematical formality in this problem, in fact, we take equivalent risks in different moments in life. Kahneman and Tversky did not model this aspect, so this paper discusses the perception of individuals about dynamics gambles.  
But specifically, the contributions of this paper consist in 
\begin{enumerate}
    \item propose a a measure of contrast between  gambles;
    \item propose an experiment, where the additive dynamics has a null contrast and nonadditive dynamics  has a high contrast;
\end{enumerate}
Thus, if individuals prospect gambles with nonadditive dynamics, they will perceive the difference between the options and a strong bias will be revealed. In order to evaluate this hypothesis,  67 psychology students participated in the experiment and two problems revealed the strong bias, surpassing 95\% of the preference for one of the options. In addition, the experiment reveals different behaviors from those predicted by Kahneman and Tversky's Prospect Theory.

\section{Different dynamics for a gamble}
When we present a gamble to an individual, and there is no communication about the replay strategy, the control of dynamics is in the individual.
Different dynamics modify the wealth composition rule and, consequently, change the gambler's reference point over time. To show this, we can consider two ways in which an individual can interpret the same gamble by taking his wealth as a point of reference, $W_0$:

\begin{itemize}
    \item {\bf gamble with additive repetition} – win a monetary amount $G$ with probability $p$ or lose a monetary amount $L$ with probability $1-p$;
    \item {\bf gamble with multiplicative repetition} – perform $1+g$ with probability $p$ or perform $1-l$ with probability $1-p$, where $g=G/W_0$ and $l=L/W_0$.
\end{itemize}

Although the dynamics above are different, we note that  the  gamble is the same in the start moment. First, the outcomes are equals. Second, the probability distribution also are equals and does not change with the repetitions. Third, the number of repetitions is limited to the individual's wealth ($0\leq L\leq W_0$ or, equivalently, $0\leq l\leq 1$).

 In gamble with additive repetition, in each trial, only an amount $G$ is added or an amount $L$ is deducted from the individual's wealth. After $n$ trials, we have:

 \[    W_n=W_0+\underbrace{G+G-L+G  -L-L +\cdots+G-L-L}_{n\textnormal{ trials}}.\]
 
 For comparison between the two dynamics, we can assume $g=G/W_0$ and $l=L/W_0$ at an instant $t_0$ (remember that $W_0$ is personal wealth at an instant $t_0$).  Thus, the average growth rate is calculated using an average weighted by the probabilities,
 \begin{equation}
         { \cal G}_0=\lim\limits_{ n\to \infty}\frac{W_n-W_0}{nW_0}=pg-(1-p)l.
           \label{simple_gamble_return_rate}
      \end{equation}
      
      On the other hand, the gamble with multiplicative repetitions has the following  development,
    \[W_n=W_0\underbrace{(1-l)(1+g)(1+g)(1-l)\cdots (1-l)(1+g)}_{n\textnormal{ trials}} .\] This leads to the average growth factor calculated by the geometric mean, \[\lim\limits_{n \to \infty}\sqrt[n]{\frac{W_n}{W_0}}= (1+g)^ p(1-l)^{1-p},\]
where we can calculate average growth rate through the continuously compounded return,
\begin{eqnarray}
   \nonumber {\cal G}_1 &=& \ln  \left[(1+g)^p(1-l)^{1-p}\right]\\
   &=&p\ln (1+g) + (1-p)\ln(1-l)
    \label{compound_gamble_return_rate}.
\end{eqnarray}

The additive and multiplicative gambles show that different dynamic models are possible for a gamble with the same odds and starting outcomes. So, change the strategy rules that define the accumulation of wealth, then the prospect on the gamble will change.  

\section{Kelly's rate and Tsallis entropy}

There is a substantial similarity between communication channels and means of exchange, such as money. A communication channel is a physical mean for exchanging messages.  Then, we measure the uncertainty between transmitter and receiver through entropy, where the symbols that leave the transmitter can undergo random changes in the communication channel, arriving modified signs at the receiver \cite{shannon1948mathematical}. Similarly, money is the means to exchange goods and services from one individual to another, where uncertainties are also present. Thus, risky monetary results can be treated mathematically as symbols of a noisy communication system \cite{kelly2011new}.

Kelly's rate is based on Shannon's entropy, but there are no limitations to develop a similar rate based on Tsallis entropy,
\[{\cal G}_q\equiv p\ln _q (1+g)+(1-p)\ln_q(1-l),\]
where
\[\ln_q \equiv \left\{\begin{array}{rc}
\frac{x^{1-q}-1}{1-q},&\mbox{if}\quad q\neq 1,\\
\ln x, &\mbox{if}\quad q=1.
\end{array}\right.
\] 

Taking $1+g=[p^{1-q} +1-q]^{\frac{1}{1-q}}$ and $1-l=[(1-p)^{1-q} +1-q]^{\frac{1}{1-q}}$ for $0<q<2$ and $p\in [1-q^{\frac{1}{1-q}},1]$, we have
\[{\cal G}_q(p)= 1 - S_q(p),\]
where $S_q(p)=-p\ln_q p - (1-p)\ln_q(1-p)$ is the Tsallis entropy. 

The growth rate modeled by nonextensive statistics can describe different dynamics of repeat gambles.
When $q=0$, we have the average growth rate of the gamble for additive repetitions,
\[{\cal G}_0= pg-(1-p)l.\]
And when $q=1$, we have the average growth rate  for multiplicative repetitions calculated through  continuously compounded rate, \[{\cal G}_1= p\ln (1+g) + (1-p)\ln(1-l).\]
Therefore, $q$ is the compound parameter, and establishes the wealth accumulation  rule over time. When $q=1$, we have a rate similar to the continuously compounded periodic interest, and the entropy is similar to the Boltzman-Gibs entropy \cite{tsallis1988possible}. 
In general, various strategies  are possible. For example, in \cite{tsallis1999nonextensive} a rule based on Polya's urn is discussed, whose expected rate of return cannot be calculated by the equations (\ref{simple_gamble_return_rate}) and (\ref{compound_gamble_return_rate}). Furthermore, generalizations of these dynamics  can be found in portfolios \cite{carr2020generalized, trindade2020portfolio} and in the intertemporal choice \cite{do2021personal}.

When comparing ${\cal G}_1$ to Kelly's work, it becomes apparent that the logarithm is calculated using the naperian base rather than the base 2. However, this difference does not pose a problem as the Boltzmann-Gibbs and Shannon entropies share logical similarities, as pointed out by E. T. Jaynes \cite{jaynes1957information}. In finance, growth rates calculated using the natural logarithm are more effective and meaningful. This can be traced back to Jacob Bernoulli's discovery of the constant $e$ in 1683 while studying compound interest \cite{o2001mactutor}. Therefore, despite differences in the logarithmic bases used in different contexts, the natural logarithm proves to be more appropriate in finance, not only due to its historical significance but also its effectiveness in determining continuously compounded growth rate.

\section{Contrast among prospects with different dynamics}
Kahneman and Tversky's experiments show how individuals have biases when options certain and uncertain are equated by weighted averages by the probabilities. This type of average is suitable for describing gambles whose repetitions are additive. To identify biases, they proposed gambles where the options did not deviate significantly from this average and checked the respondents' preferences, where most of their experiments isolate gains and losses. 

We can also evaluate prospects from a dynamic perspective, also isolating gains and losses. First let us evaluate gains only. Consider that two gain options are presented to an individual, where only one of them can be choose. In order to make these options equivalent with additive repetitions, we must define  
\begin{enumerate}[label=]
\item $A)$  to win $Mp$;
\item $B)$  to win $M$ with probability $p$.
\end{enumerate}
Both options have equivalent expectation when outcomes are compared by weighted average by the probabilities, i.e, the expectation is $Mp$ in both options.

Similarly, risky losses can also be defined. If we replace the word ``win'' for ``lose'' in the options $A$ and $B$, then we have the following gambles that result in the wealth decreasing:  
\begin{enumerate}[label= ]
\item $A)$ to lose $Mp$;
\item $B)$ to lose $M$ with probability $p$.
\end{enumerate}
In this case, the weighted average by the probabilities is $-Mp$ in both options.

Now we must estimate the behavior of risk takers assuming different dynamics in the above gambles. First, we must note that uncertain outcomes have probability $p$ while certain outcomes are weighted by probabilities. Thus, if an individual evaluate the gambles  by additive repetitions, both options are equivalents. In this assumption, there is not contrast between the options $A$ and $B$.  However, if there is a expectation of  repetitions nonadditive,   then the prospects are different, and the difference will be perceive if the contrast is high. Like this, we can calculate the contrast between options $A$ and $B$ by
\begin{equation}
    C=\left|20\log\left(\frac{p_A\ln_{q} (1+g_A)+(1-p_A)\ln_{q}(1-l_A)}{p_B\ln _{q} (1+g_B)+(1-p_B)\ln_{q}(1-l_B)}\right)\right|,
    \label{eq:contrast}
\end{equation}
where $20\log$ establishes a scale of contrast in dB, $p_A$ and $p_B$ are the probabilities of gain, and 
\begin{eqnarray}
\nonumber g_A  &=& G_A/W_0=Mp/W_0,  \\
\nonumber g_B  &=& G_B/W_0=M/W_0, \\
\nonumber l_A  &=& L_A/W_0=-Mp/W_0,  \\
\nonumber l_B  &=& L_B/W_0=-M/W_0
\end{eqnarray}
 are relations between the outcomes and the reference point (individual wealth). 
 
Assuming $x$ represents the outcome over wealth,  we can consider that $x\in [-1,\infty[$. There are two main reasons for this type of normalization. Firstly, the losses are inherently limited by the individual's wealth. Secondly, the potential gains can be unlimited. Thus, the contrast to compare uncertain outcomes with  certain outcomes have the following form
\begin{equation}
    C=\left|20\log\left(\frac{\ln_{q} (1+px)}{p\ln _{q} (1+x)}\right)\right|,
    \label{eq:contrast1}
\end{equation}
where $\ln_{q} (1+px)$ is the growth rate of certain outcome, and  $p\ln _{q} (1+x)$ is the growth rate of uncertain outcome.

\begin{figure}[ht]
    \centering
    \includegraphics[scale=0.7]{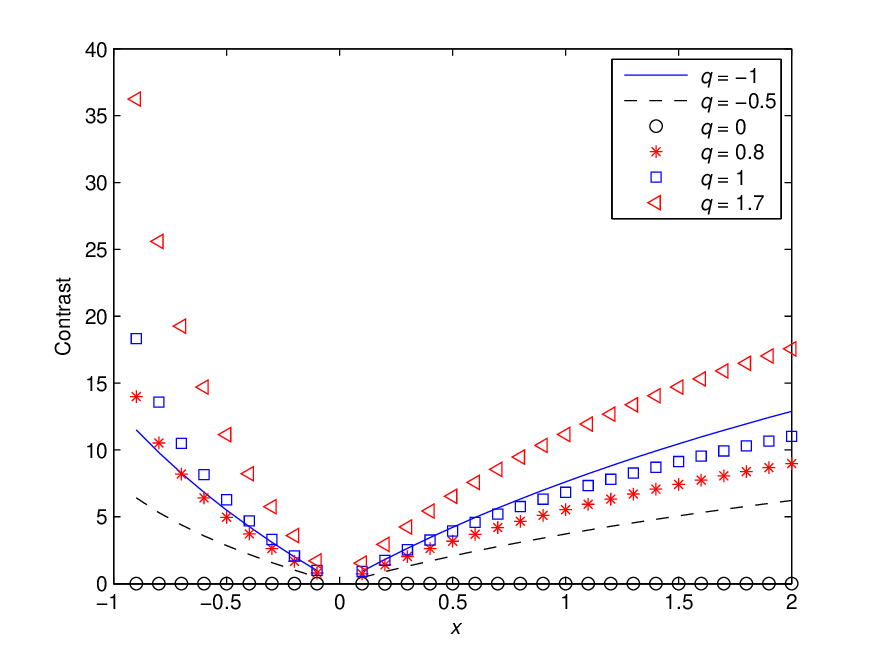}
    \caption{Contrast versus  $x$  for some values of $q$ and $p=0.05$.}
    \label{fig:contrast}
\end{figure}

Figure \ref{fig:contrast} allows us to understand the contrast between options $A$ and $B$   varying $x$ for some values of  $q$: 
\begin{enumerate}
    \item In $q=0$ the contrast between the options is null because the $A$ and $B$ propose equivalence for gambles with additive repetitions (see black ball horizontal line, where the contrast is null for all $x \in [-1,2]$);
    \item the contrast is stronger for negative prospects ($x<0$) than positive prospects ($x>0$);
    \item the contrast is concave for positive prospects and convex for negative prospects; 
    \item all dynamics have low contrast when the outcomes are much smaller than the reference point ($C\approx 0$ when $x\approx 0$); 
\end{enumerate}

 If all individual adopted  additive repetitions in their strategies, they should be indifferent between $A$ and $B$ options. But Kahneman and Tversky showed that such indifference does not exist.  However, their experiments are not   far from the reference point that they identified, what makes it difficult to identify other dynamics because the contrast is low. 
 
  The contrast is relevant when $x$ moves away from zero, where other dynamics can produce strong contrasts (see Figure \ref{fig:contrast} for  $q \neq 0$).  Therefore, if such contrast exist, then this will strongly bias risk takers' choices. To verify this hypothesis, it is necessary purposefully significant outcomes regarding the wealth of individuals.

\section{Experimental results}

Now, we need see if people facing gambles have  prospects for nonadditive repetitions. To verify this hypothesis, respondents answer questionnaires that presented risky situations, where
\begin{itemize}
    \item the options $A$ and $B$ in each problem have expectation calculated by an average weighted by the probabilities. Thus, case individuals prospect values close to averages with additive dynamics, then the contrast will be low and choices will be balanced (approximately fifty percent for each option);
    \item otherwise, if individual prospect through nonadditive dynamics, then the contrast will be high because the outcomes are close or beyond to the reference point. Thus, the perception of individuals identify the high contrast  and will reveal a strong bias in the results.
\end{itemize}
In the experiment, all respondents are psychology students from the Universidade Federal do Cear\'{a}, Campus Sobral, so it is assumed that they do not have any training in time averages. A total of 67 students answered the questionnaire and the percentage of choice in each option is in brackets. For a comparison between the contrast estimate and the students' response, Figure 2 presents a contrast estimate for each problem using equation \ref{eq:contrast}.

\begin{figure}[ht]
    \centering
    \includegraphics[scale=0.7]{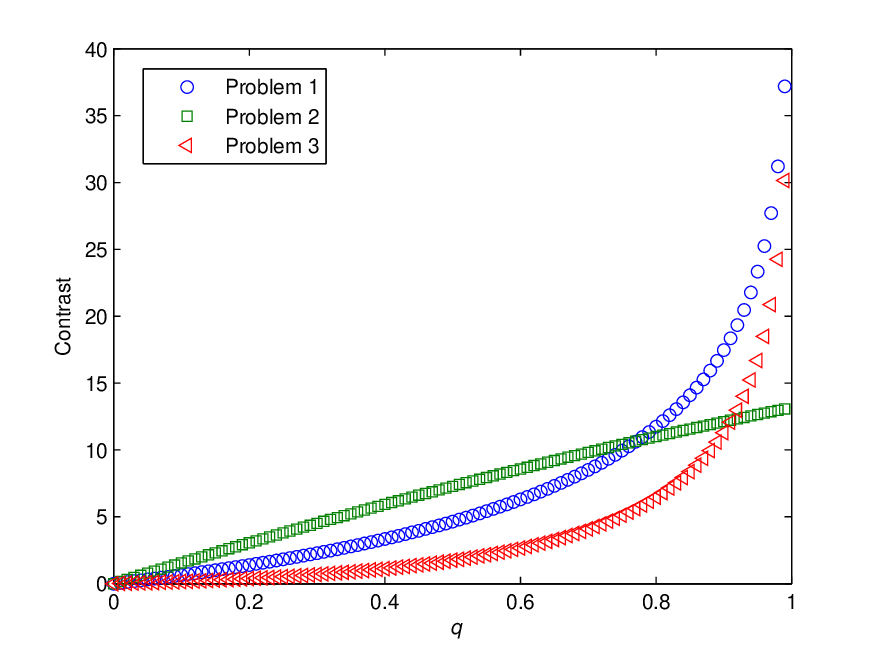}
    \caption{Contrast versus  compound parameter ($q$) -- 
        all contrasts are null in $q=0$, where the dynamic is additive. However, the contrasts   grow until $q=1$, where the dynamic is multiplicative. 
        In Problem 1, all losses are proportionally compared with the reference point. Then, the data for calculating the contrast are: $p_A=1/2$; $g_A=0$; $l_A=1$; $p_B=1/2$; $g_B=0$; $l_B=1/2$.  In Problem 2, we know that $g_A=20g_B$. However, there are no statistical data to estimate $g_B$. Thus, $g_B=0.5$ is a chosen value  just to facilitate the visualization of the curves, as this value does not prejudice  the conclusions because it is the only one  concave curve in the graph. Like this, the other data for calculating the contrast are: $p_A=0.05$;  $l_A=0$; $p_B=1$;  $l_B=0$. 
        In Problem 3, the reference point is shifted with the assumption of receiving an inheritance much greater than the wealth of individuals. Then, two options that put the heritage at risk are presented. Thus, the data for calculating the contrast are: $p_A=0.05$; $g_A=0$; $l_A=1$; $p_B=0$; $g_B=0$; $l_B=190/200$.  }
    \label{fig:contrastP1P2P3}
\end{figure}

Problem 12 presented in \cite{kai1979prospect} consists of choosing between an uncertain loss with maximum entropy and a certain loss. In this case, Kahneman and Tversky found that most individuals prefer a risky loss rather than a certain loss. This result is known as strong evidence that individuals take risks when they are into a loss context. However, if we increase the losses magnitudes, the contrast increase and the risk seeking is abandoned. The problem below shows how people present a risk aversion in situations of losses that can lead an individual to ruin.
\\
\\
{\bf Problem 1} -- Imagine all the material possessions you own: car, home, bike, clothes, money, etc. If you have a big debt and they are offering the following payment options, then which do you prefer? 
\begin{enumerate}[label=\alph*)]
    \item Toss a coin. If it comes up heads, you loss all your goods, and if it comes up tails, you loss nothing; [1.49\%]
    \item to lose half of all you own. [98.51\%]
\end{enumerate}
A total of 66 individuals out of 67 preferred to lose half of all  their goods, rather than risking losing everything after the coin toss.
In this experiment, it is clear that there is a very strong contrast between the options, as almost all respondents demonstrated to be prudent in this context. In conversation with some respondents outside this sample, arguments such as we ``it is easier to continue with half than to start over'' were presented. This type of argument shows that individuals see gambles as dynamic processes and have the time as a determining factor in their choices.

In the Prospect Theory, one of the fourfold patterns shows that most individuals prefer small chances of receiving a premium than equivalent amounts weighted by probabilities. In everyday life, this phenomenon is similar to buying  raffle tickets \cite{kahneman2011thinking}). In the experiment below, we can also see that this behavior is not general.
\\
\\
{\bf Problem 2} -- Imagine that two opportunities are offered to you to become a millionaire person and you can only choose one. So, which do you prefer?
\begin{enumerate}[label=\alph*)]
    \item to win  200 million dollars with probability 0.05; [4.48\%]
    \item to win 10 million dollars with certainty. [95.52\%]
\end{enumerate}
Only 4 individuals out of 67 expressed interest in taking risks. So, what's the difference between 1) rejecting \$ 10 to stay exposed to winning \$ 200 with 5\% chance, and 2) rejecting \$ 10 million to stay exposed to winning \$ 200 million with the same chance? When we increase the rewards to close (or beyond) reference point, then their impact on the individual's wealth changes.  Arguing about time, \$ 10 million is an amount that takes a long time to acquire by most individuals, and an opportunity to acquire it quickly is hard to miss. Therefore, the risk aversion manifested by respondents facing high amounts is consistent with a nonadditive prospect.
 
 Another behavior of the fourfold pattern is risk seeking even when the probability of loss is high. An analogous problem with very high losses  can be seen below:
\\
\\
{ \bf Problem 3} - Imagine you received an inheritance equivalent to \$ 200 millions. However, a problem has arisen upon receipt of inheritance  and you will have to choose between:
\begin{enumerate}[label=\alph*)]
    \item 95\% chance to lose the \$ 200  million; [26.87\%]
    \item to pay \$ 190 million. [73.13\%]
\end{enumerate}
In Kahnemam and Tversky's experiments without extreme outcomes, risk seeking is preferable even when the probability of loss is very high \cite{tversky1992advances}, but in the above problem, this phenomenon is not observed. First, respondents do not seem to ignore inheritance what shifts their point of reference. Second, they demonstrated risk aversion, because most sought the certain loss that still guarantees a fortune of \$ 10 million (a hard amount to achieve over time). So, one of the quadruple pattern, which suggests risk seeking even when the probability of loss is high, can be violated when the reference point is shifted.

Finally, it should be noted that the estimated contrasts were consistent with the observed behaviors. First, problems 1 and 2 presented the strongest biases of the experiment (above 95\% of preferences) corroborating with high contrasts (see Figure \ref{fig:contrastP1P2P3}). Second, Problem 3 has a weaker contrast for most of the range,  $q\in [0,0.9]$,  because the curve is more convex (see Figure \ref{fig:contrastP1P2P3}).  This is consistent with the low unanimity in the preference for a certain loss (option with only 73.13\% of preferences).

\section{Conclusion}
Do risk takers prospect nonadditive dynamics? This paper proposes a method to measure the contrast between gambles and presents an experiment where the options have high contrast when nonadditive dynamics are in the strategy. Thus, if individuals adopt nonadditive repetitions, they must perceive the difference and reveal a strong bias. A total of 67 psychology students answered a questionnaire, and the results proved the methodology's effectiveness because two problems showed strong bias (preferences above 95\%). Furthermore, predicted behaviors by Prospect Theory, such as risk seeking for negative prospects and two fourfold patterns, are violated.

Finally, we can conclude that the repetition dynamics is an essential physical aspect in gambles, and it strongly affects the behavior of people. Therefore, including dynamics gambles in psychophysical models will allow more accurate diagnoses for a wider range of risky situations.

\bibliographystyle{plain}
\bibliography{ref}
\end{document}